# Interpretable graph-based models on multimodal biomedical data integration: A technical review and benchmarking

Alireza Sadeghi[1], Farshid Hajati*[2], Ahmadreza Argha[3], Nigel H Lovell[3,6], Min Yang[4], Hamid Alinejad-Rokny*[5]

[1] Holcombe Department of Electrical and Computer Engineering, Clemson University, Clemson, South Carolina, USA

[2] School of Science and Technology, Faculty of Science, Agriculture, Business and Law, University of New England, Armidale, NSW, 2350, Australia

[3]School of Biomedical Engineering, UNSW Sydney, Sydney, NSW, 2052, Australia.

[4]Shenzhen Institute of Advanced Technology, Chinese Academy of Sciences, Shenzhen, China.

[5] UNSW BioMedical Machine Learning Lab (BML), School of Biomedical Engineering, UNSW Sydney, Sydney, NSW, 2052, Australia.

[6] Tyree Foundation Institute of Health Engineering (IHealthE), UNSW Sydney, Sydney, NSW, 2052, Australia.

* To whom correspondence should be addressed to H.A.R. (h.alinejad@unsw.edu.au)

## Abstract

In healthcare, integrating diverse biomedical data modalities is critical for comprehensive analyses that can lead to better patient outcomes. Graph-based models have emerged as effective tools for handling such complex, non-Euclidean data by leveraging spatial and relational structures. However, model interpretability remains crucial for their successful deployment and for regulatory-approved implementations in clinical settings, as healthcare practitioners need transparent insights to build trust and guide decision-making. In this work, we present a technical and comprehensive review of research on interpretable graph-based models applied to multimodal biomedical data, covering studies from January 2019 to September 2024. Our analysis of published papers reveals a dominant focus on disease classification (particularly cancer) and a heavy reliance on static graph construction methods (e.g., Pearson correlation, Euclidean distance). While some approaches incorporate inherently interpretable designs, most rely on post-hoc techniques originally developed for non-graph data—such as gradient-based saliency or SHAP (Shapley additive explanations)—and only a small fraction utilize graph-specific methods like GNNExplainer. To address these gaps, we categorize existing explainable artificial intelligence (XAI) approaches according to their interpretability strategies and highlight evolving trends, including the graph-in-graph paradigm, the integration of knowledge graphs, and dynamic edge construction. To illuminate their practical implications, we compare and benchmark representative XAI techniques (sensitivity analysis, gradient saliency, SHAP, and graph masking) using a real-world Alzheimer's disease (AD) dataset. Our findings show that while SHAP and sensitivity analysis reveal a broader spectrum of established disease-related pathways and Gene Ontology terms, gradient saliency and graph masking highlight unique metabolic and transport processes, underscoring the complementary strengths of these methods. To foster broader adoption, we provide a step-by-step flowchart for developing interpretable graph-based models, offering guidance on key architectural choices and optimization strategies. Taken together, this review not only summarizes the current state of interpretable graph-based modeling for multimodal medical data but also outlines emerging directions—such as integrating new explainability methods and expanding to underexplored diseases—thereby serving as a vital reference for researchers and practitioners in this rapidly evolving field.

## 1. Introduction

Deep learning (DL) has emerged as a transformative technology across various fields, primarily due to its unmatched ability to process vast amounts of data and uncover complex patterns [1, 2]. In medical research, the impact of DL is particularly profound. Its capacity to analyze complex medical data such as images, genomic sequences, and patient records has transformed diagnostics [3, 4], treatment planning [5, 6], and even drug discovery [7, 8]. Moreover, DL's predictive capabilities are driving the development of personalized medicine by enabling the assessment of individual patient risks and responses to treatments [9]. However, traditional DL models like neural networks (NNs) and convolutional neural networks (CNNs) often focus on homogeneous data, overlooking crucial relationships across different modalities. This limitation can hinder their effectiveness in scenarios requiring the integration of diverse data types—such as medical images, clinical notes, and genomic data—for more accurate diagnoses and personalized treatments [10-12]. By overcoming these challenges, DL continues to advance healthcare, improving patient outcomes and quality of life.

Unlike traditional DL, graph-based models offer a robust solution to these challenges by effectively handling non-grid structured data and naturally representing complex relationships through nodes and edges in a graph structure. These models excel at integrating heterogeneous data sources, capturing both intra-modality and inter-modality interactions, and uncovering nuanced patterns that enhance performance in multimodal medical data analysis tasks. Their adaptability makes them particularly suitable for complex healthcare applications requiring comprehensive data integration and analysis.

Despite the promising capabilities of both DL and graph-based models in genomics [13, 14], their widespread adoption in critical healthcare settings is impeded by their "black-box" nature, which provides limited transparency into their decision-making processes [15]. This lack of interpretability raises concerns regarding trust, accountability, and ethical use, underscoring the necessity for models that are not only accurate but also interpretable. Enhancing model interpretability enables clinicians and researchers to understand, validate, and trust the predictions, thereby facilitating informed decision-making and fostering confidence in deploying these technologies in sensitive medical contexts. Interpretability and explainability is also critical for minimizing the barriers to translation of such algorithms into regulatory-approved medical systems.

While some work has already begun exploring interpretable graph-based models for analyzing multimodal medical data, there remains significant scope for future research to further advance this field. Recognizing this emerging field, this paper presents a technical review of interpretable graph-based models applied to multimodal medical data. We aim to provide comprehensive insights and guidance for researchers and practitioners interested in this emerging field. To the best of our knowledge, this review represents the first technical exploration at the intersection of interpretability, graph-based modeling, and multimodal medical data analysis, setting the stage for future advancements in this crucial area. The main contributions of this study are:

- We systematically review existing studies that employ graph-based models on multimodal medical data with integrated interpretability techniques, examining their objectives, utilized modalities, graph construction methods, and approaches to achieving interpretability.

- We also categorize and benchmark various interpretability methods applied to graph-based models, providing a structured understanding that can aid researchers in selecting appropriate techniques for their specific applications.

- Recognizing scenarios where inherently interpretable models are preferable despite potential performance trade-offs, we present a step-by-step flowchart to guide the development of such models in future studies.

In the rest of this paper, we begin with the Methodology section outlining our approach for retrieving and selecting relevant literature. The Overview of Common Graph-Based Models on Multimodal Biomedical Data section provides a concise description of prevalent graph-based architectures used in this context. In the Graph Construction in Reviewed Studies section, we explore various strategies employed to construct graphs from diverse medical data sources. The Interpretation in Multimodal Biomedical Data Analysis section discusses the interpretability techniques used to enhance model transparency and trustworthiness. Finally, we present our findings and recommendations in the Discussion section and conclude the paper in the **Conclusion** section.

## 2. Methodology

To gather relevant research papers, we used three primary sources: Google Scholar, Scopus, and PubMed. We conducted searches using various combinations of the following keywords related to our study:

- Interpretable, explainable, or transparent graph networks
- Multimodal or integrated

- Medical, biomedical, biological, healthcare, genetic, or genomic data

This comprehensive search yielded a total of 2,786 papers published between January 1, 2019, and September 31, 2024. Subsequently, we screened these papers by their titles, leading to the inclusion of 350 papers for further examination. The abstracts of these selected papers were reviewed in the next step, resulting in the identification of 124 papers for a more in-depth analysis. Upon analyzing the full-length papers, 98 of them were excluded for various reasons. Some papers were duplicates, others used graph-based models in different contexts unrelated to multimodal data analysis, and some failed to provide a clear explanation of the interpretability techniques employed in their models. This filtering process left us with 26 papers eligible for final review and inclusion in our study. To illustrate our selection process, we have depicted a PRISMA diagram in **Figure 1**, outlining the stages of paper selection and exclusion. It is important to note that each step of the methodology described was conducted independently by all authors involved in this study.

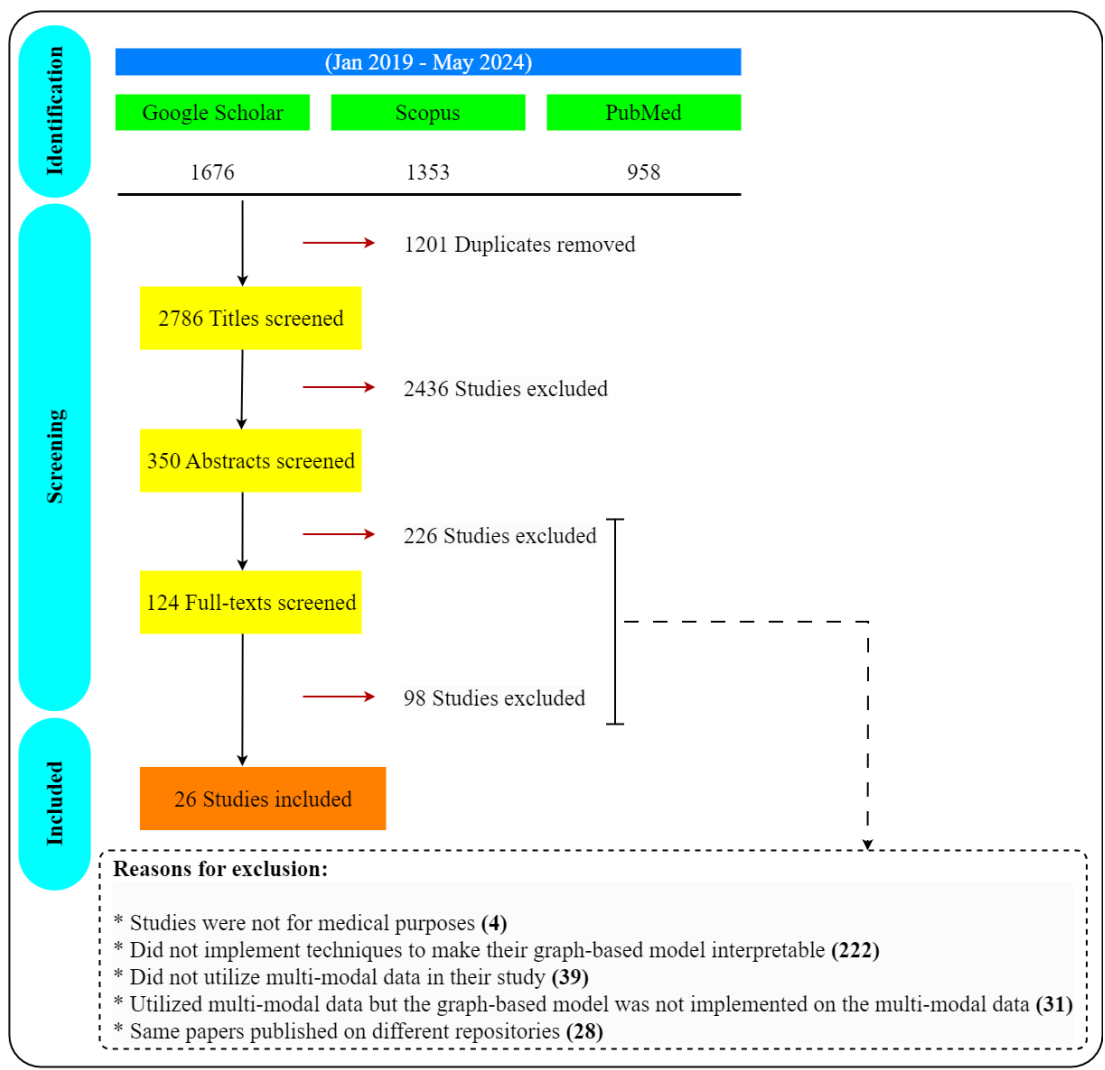

**Figure 1. The PRISMA diagram illustrates the process of retrieving related papers for this review study**.

## 3. Literature overview

Graph-based models have become integral in analyzing multimodal biomedical data for various regression and classification tasks related to cancer analysis, disease diagnosis, and biomolecular interactions. **Table 1** provides an overview of these models, including deep neural networks (DNNs), graph neural networks (GNNs), graph isomorphism networks (GINs) [16], graph convolutional networks (GCNs), graph attention networks (GATs) [17], and graph transformer networks (GTNs) [18], along with their key characteristics. **Figure 2a** illustrates the architectural schematic of these models, while **Figure 2b** displays their distribution across surveyed studies. These models are designed primarily to accomplish two main objectives: performing graph-level tasks or node-level tasks (refer to **Figures 2c** and **2d**). In graph-level tasks, each sample in the raw dataset is represented as a graph.

**Table 1.** The summary of various graph-based models employed in the surveyed studies alongside their respective formulas.

| Model | Explanation | Formula | Denotation |
|---|---|---|---|
| DNN | In graph-related studies, a feedforward framework built using the nodes and edges of the graph as input. | $z_n = \phi_n (W_n X_n + b_n)$ | $z_n$: The output of the neuron $n$. $\phi_n$: The activation function applied to the neuron $n$. $X_n$: The input to the neuron $n$. $W_n$ and $b_n$: The weights and bias inputted to the neuron $n$. |
| GNN | A framework for updating each node's features by aggregating the features of its neighboring nodes | $h_i^{l+1} = COMB^l(h_i^l, AGG^l (\{h_j^l : j \in \mathcal{N}_i\}))$ | $h_i^{l+1}$ and $h_i^l$: Node $i$'s feature vector in layer $l+1$ and $l$, respectively. $COMB^l$: The combination function in layer $l$. $AGG^l$: The aggregation function in layer $l$. $\mathcal{N}_i$: Node $i$'th neighborhood. |
| GIN | An architecture with adjustments to the GNN, to make it more powerful | $h_i^{l+1} = MLP^l((1 + \epsilon^l) \cdot h_i^l + \sum_{j \in \mathcal{N}_i} h_j^l)$ | $h_i^{l+1}$ and $h_i^l$: Node $i$'s feature vector in layer $l+1$ and $l$, respectively. $MLP^l$: a multi-layer perceptron in layer $l$. $\epsilon$: a learnable parameter or a fixed scalar. $\mathcal{N}_i$: Node $i$'th neighborhood. |
| GCN | A framework for updating the features of graph nodes through convolutional filtering applied to the input graph signals | $H^{l+1} = \phi^l (AH^l \theta^l)$ | $H^{l+1}$ and $H^l$: Nodes' feature vector in layer $l+1$ and $l$, respectively. $\phi^l$: Non-linear activation function of layer $l$. $A$: The adjacency matrix. $\theta^l$: Layer $l$'s convolutional filter's weight matrix. |
| GAT | An attention-based architecture for computing the hidden representation of each node by considering the | $\alpha_{ij} = softmax_j (a (Wh_i, Wh_j))$ | $\alpha_{ij}$: Attention coefficient of node $j$'s features to node $i$. $a$: The shared attention mechanism (single-layer feedforward NN). |
| | | importance of features from other nodes | | $h_i$: Node $i$'s feature vector. $W$: weight matrix of the shared linear transformation |
| GTN | A framework similar to natural language processing transformers but made for graph inputs | $h_i^{l+1} = O_h^l \left( \Big\|_{k=1}^{H} \sum_{j \in \mathcal{N}_i} (w_{i,j}^{k,l} V^{k,l} h_j^l) \right)$ | $h_i^{l+1}$ and $h_i^l$: Node $i$'s feature vector in layer $l+1$ and $l$, respectively. $O_h^l$: An additional linear transformation in layer $l$. $\|$: Concatenation operation. $k$: The attention head (from 1 to $H$ heads). $w_{i,j}^{k,l}$: The weight of the $k$'th attention head calculated by features from node $i$ and $j$ in layer $l$. $V^{k,l}$: The Value vector of node $i$ in the $k$'th attention head in layer $l$. $\mathcal{N}_i$: Node $i$'th neighborhood |

Labeled graphs, derived from labeled samples, are used to train the graph-based model. Subsequently, this trained model is employed to make predictions for unlabeled graphs (samples). On the other hand, in node-level tasks, each sample is represented as a node within the constructed graph. The objective is to generate predictions for unlabeled nodes by leveraging information from the labeled nodes in the graph. This approach focuses on predicting individual nodes rather than the entire graph structure.

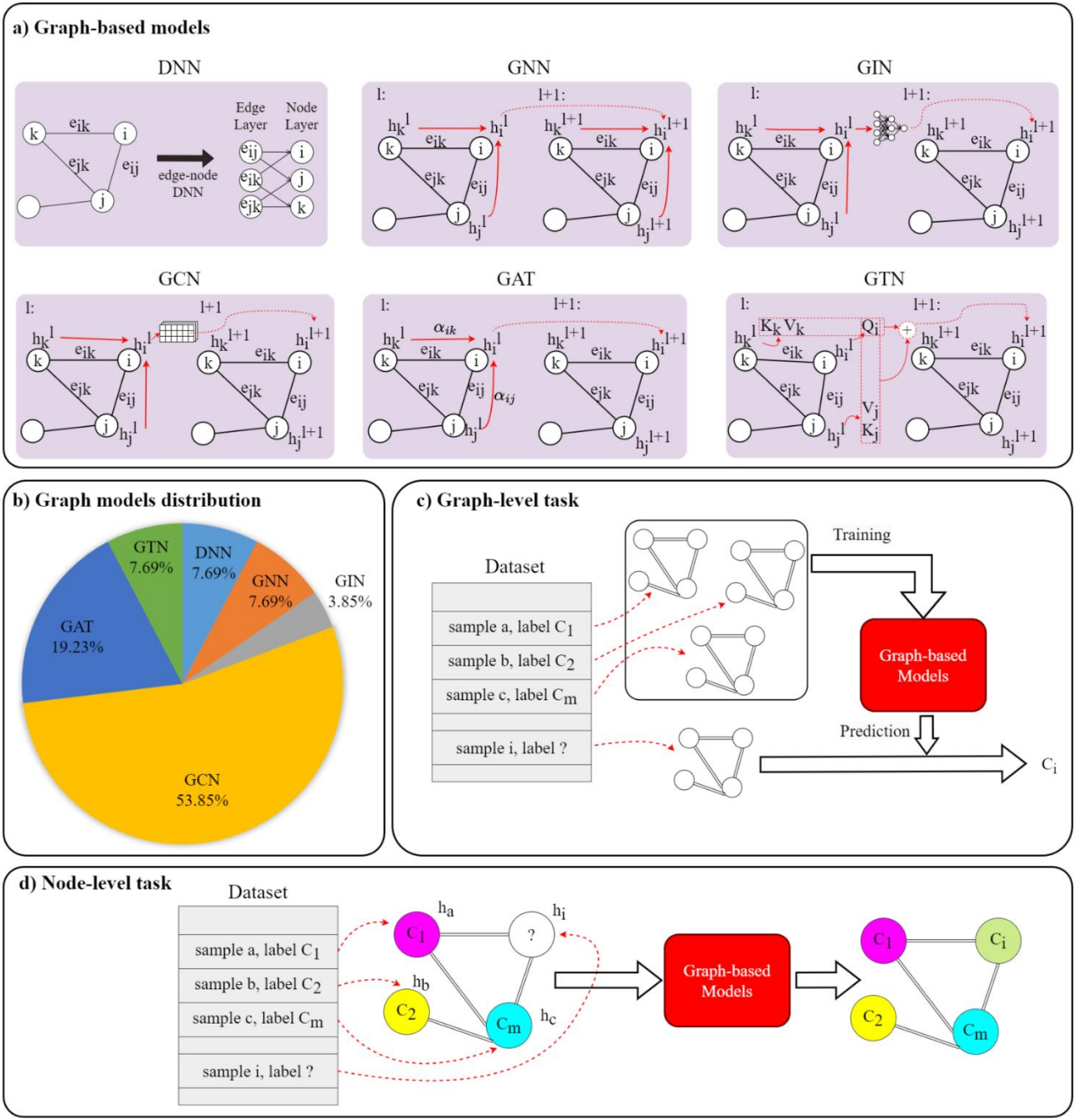

**Figure 2. An overview of graph-based models. a)** Architecture showcasing various graph-based models utilized in studies on multimodal data. **b)** Distribution of different graph-based models in the reviewed studies (n=26). **c)** The process of implementing graph-based models for graph-level analysis. **d)** The process of implementing graph-based models for node-level analysis. DNN: deep neural network, GNN: graph neural network, GIN: graph isomorphism network, GCN: graph convolutional network, GAT: graph attention network, GTN: graph transformer network.

## 4. Graph construction in the reviewed studies

The graph-based models discussed in the previous section rely on graph-structured data, making the creation of a graph from the raw dataset a critical aspect of the studies reviewed.

The graph input is typically represented as $\mathcal{G} = (\mathcal{V}, \mathcal{E})$, where $\mathcal{V}$ denotes the nodes of the graph and $\mathcal{E}$ represents the edges that link interacting nodes. Nodes are associated with their own feature vector $\mathcal{F}(\mathcal{V})$, usually derived from the various data modalities. Throughout the training phase, each node undergoes updates based solely on the information exchanged with its neighboring nodes. This mechanism ensures that only data with similar characteristics, which interact with each other, contribute to the final prediction generation.

The objective of a study and the modalities it employs play a crucial role in defining how nodes are constructed within the resulting graph. For example, in research focused on brain imaging, brain regions of interest (ROI) are commonly designated as nodes [19-26]. Conversely, studies analyzing cancers at a genomic level [25, 27-33], or those involved in drug repurposing [34], often utilize genes and proteins as nodes in their graphs. In broader disease prediction studies, such as COVID-19 or Alzheimer's classification, each node typically represents a patient, making these studies node-level analyses [35-43]. In a study by Tang et al. [44], aimed at predicting hospital readmission, each hospital admission is treated as a node. The feature vectors of the nodes across all the reviewed studies are derived by processing the specific modalities employed in each study.

To establish edges between nodes, most studies employ the static method, wherein edges are created between pairs of nodes, and the constructed graph remains unchanged throughout the training process. Often, these edges are determined based on domain-specific background knowledge. For example, several studies [27-30, 33, 34] construct edges using known interactions such as protein-protein, protein-drug, and gene-gene relationships. Another commonly used technique involves calculating the similarity between node feature vectors using different similarity functions. Edges are then maintained between nodes with higher similarity while others are either dropped using a specified threshold or the K-nearest neighbor technique. In the former method, a threshold is defined, and edges with similarity values lower than this threshold are discarded. In the latter technique, each node is connected to its K neighbors with the highest similarity. In determining similarity functions, most studies utilize Pearson correlation [19, 20, 22, 25, 37], Euclidean distance [23, 38, 41, 44], and cosine similarity [36, 42] in their research. Additionally, some studies employ other task-specific methods tailored to their particular field of study. For example, Kan et al. [26] constructed a structural network using dMRI images, where the nodes were connected based on the structural connectivity probabilities calculated using dMRI tractography algorithm.

While many studies rely on static graph construction methods, some researchers have adopted dynamic approaches. For instance, Kazi et al. [40] calculated the Euclidean distance between nodes and retained edges based on a dynamically defined threshold that adjusts throughout the training process. Similarly, Ouyang et al. [43] employed weighted cosine similarity between nodes, with weights iteratively learned during training. These researchers contend that constructing the graph structure is a crucial aspect of graph processing tasks, asserting that static graphs may be suboptimal and can lead to diminished performance. Dynamic graph construction addresses this issue by allowing the graph's topology to evolve during the training, thereby offering the potential for improved performance and better adaptation to complex relationships within the data.

Overall, graph construction is a crucial aspect of graph-based models applied to multimodal medical data. The methods for constructing these graphs vary depending on the specific study and the types of data modalities involved. Generally, nodes, node features, and edges are defined based on the unique characteristics of the data under analysis. Most studies rely on static graphs, where the connections between nodes remain fixed over time. Popular methods for creating edges between nodes include Pearson correlation, Euclidean distance, and cosine similarity. **Figure 3** presents an overview of the methods employed by different studies to construct graphs in their research, along with distributions of the techniques utilized.

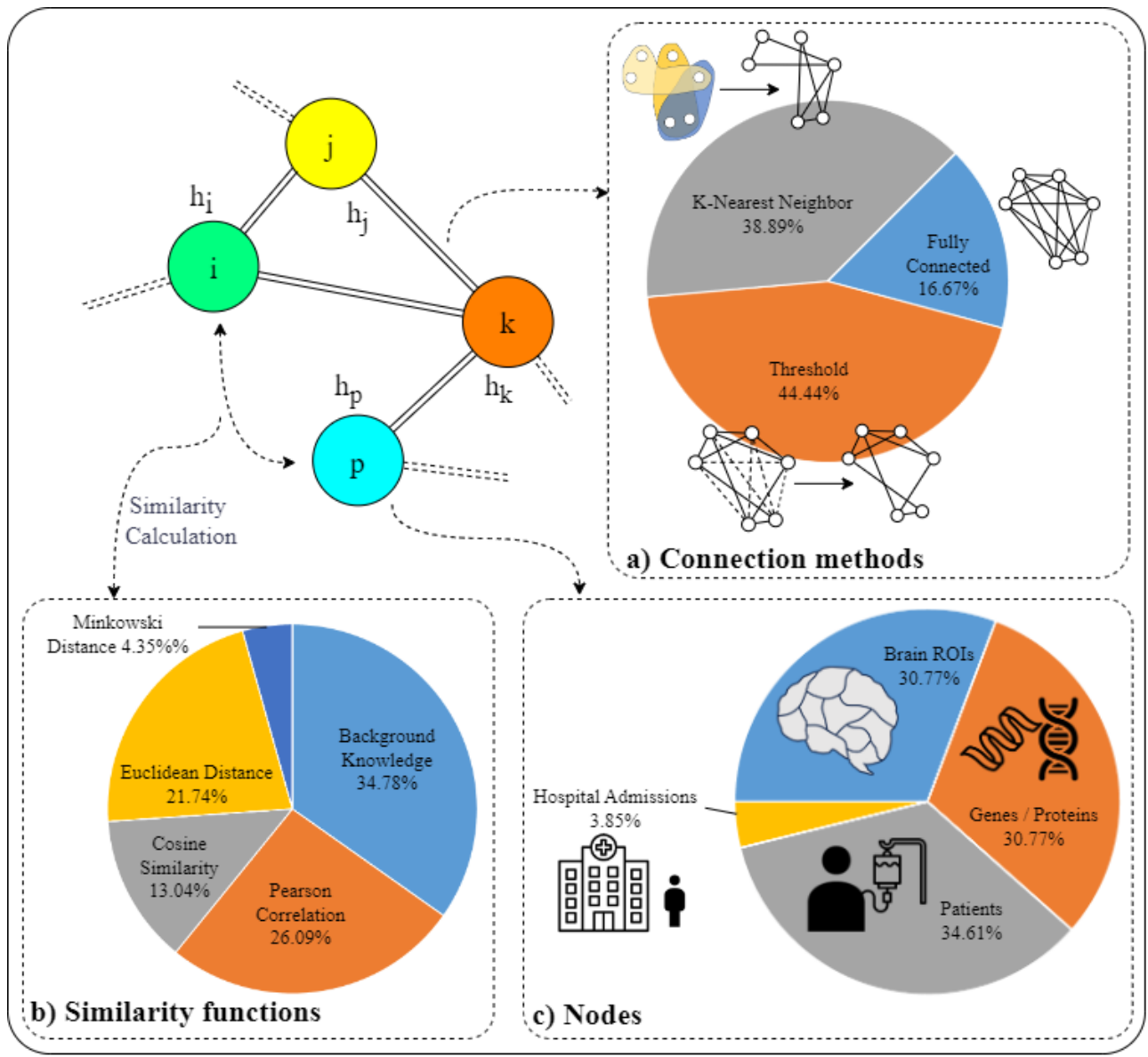

**Figure 3. Overview of various techniques employed in different studies to construct graphs for their research**. **a)** Distribution of various connection methodologies in graph

construction utilized by surveyed studies. **b)** Distribution of various similarity functions used to draw edges between nodes in the constructed graphs. **c)** Distribution of different entities implemented as nodes in the constructed graphs.

## 5. Interpretation in multimodal biomedical data integration

DL models, including graph-based ones, are often viewed as black-box systems, which poses a significant challenge when deploying them in trust-sensitive fields like healthcare. To tackle this challenge, researchers in the surveyed studies have explored various approaches to make these models explainable. We examined all the methods implemented in existing studies and categorized them into the following four categories:

- Category I: Modality or Feature Elimination
- Category II: Non-Graph XAI
- Category III: Graph-Based XAI
- Category IV: Inherent Interpretability

In the following subsections, we explore each category in detail, discussing their specific use cases. For the fourth category, which focuses on developing inherently interpretable graph-based models, we provide a step-by-step guide to assist researchers in creating transparent graph-based models in their studies. Finally, we conclude this section with a benchmark analysis, where we use methods from each category on a specific graph-based medical study to practically evaluate strengths and weaknesses of each interpretability category.

### 5.1. Category I: Modality or Feature Elimination

In the context of graph-based models applied to multimodal medical data, a primary focus for researchers is determining the significance of each modality. They seek to understand whether the utilization of multiple modalities enhances their model's performance compared to scenarios involving a single modality, and if so, to what extent. A common technique employed by most of the reviewed studies involves removing a specific modality and evaluating the model's performance without the features from that specific modality (**Figure 4**). The degree of performance decline indicates the importance of that modality to the model's overall performance. For instance, Yang et al. [19] substituted the features from one modality with dummy variables and evaluated the model's performance, while other reviewed studies opted to eliminate a certain modality, retrain the model, and recalculate its performance to demonstrate the significance of that particular modality.

However, modality elimination provides only a basic insight into the importance of each modality. For a comprehensive analysis, it is crucial to evaluate the model across all possible combinations of modalities to demonstrate the significance of that particular modality. Moreover, this method serves as an initial step in several non-graph XAI techniques like SHAP [45]. Despite some similarities, there are key differences that set them apart. In modality elimination, one modality is removed, and the model is retrained from scratch on the remaining modalities. This retraining process, which focuses exclusively on the remaining features, is not employed in SHAP and similar methods. Therefore, modality elimination can be considered as a distinct category.

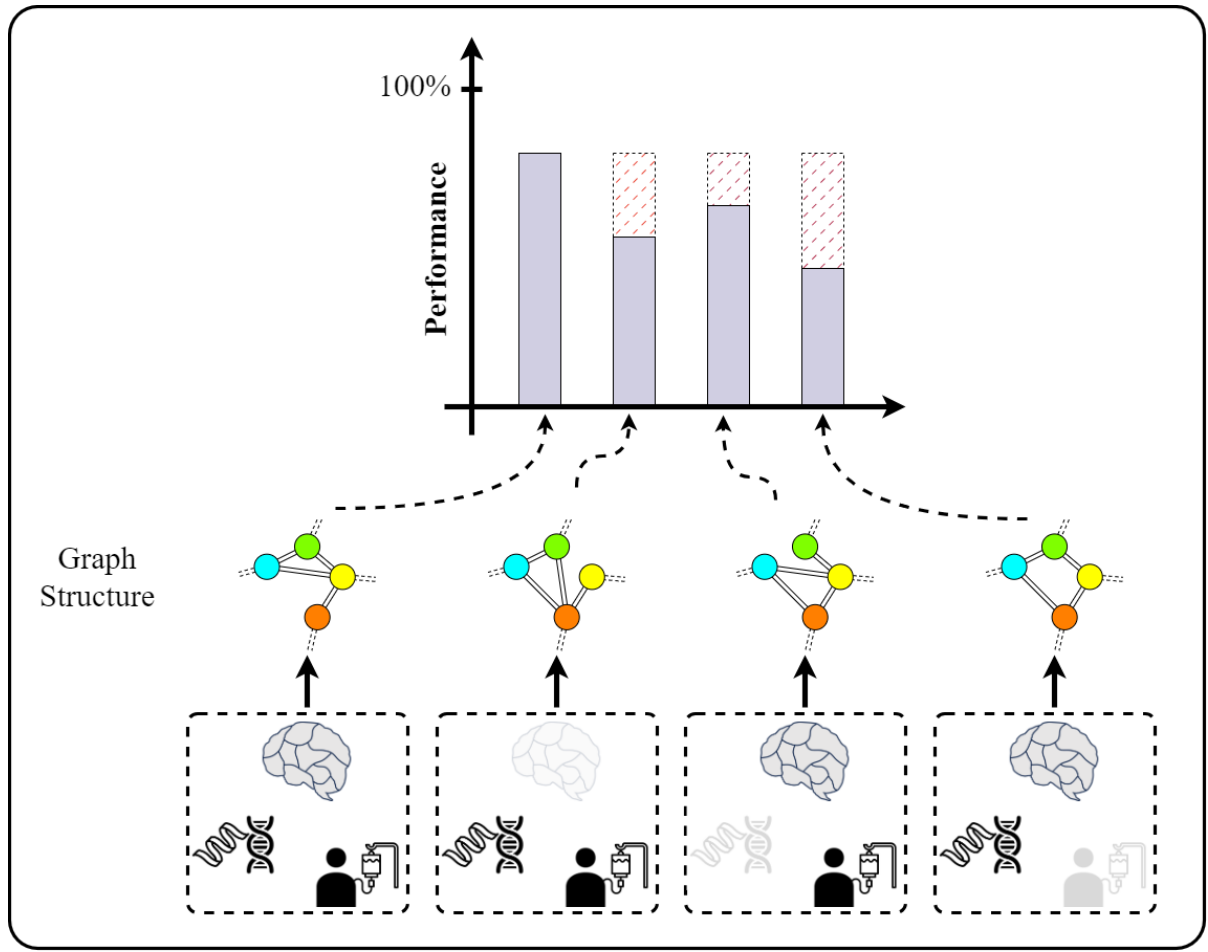

**Figure 4. Implementation of the modality elimination technique to assess the importance of each modality in a graph-based model.** In this technique, one modality is removed at each step and the model is retrained using the remaining modalities. This process allows for the calculation of the model's performance without the influence of the eliminated modalities. By comparing the decline in performance for each modality elimination, insights into the importance of each modality are obtained. Note that eliminating a modality may alter the structure of the graph-based model.

Modality elimination can introduce a limitation by potentially altering the graph structure. This alteration is particularly significant when the graph structure relies on the similarity between node feature vectors. Such changes can skew the analysis of modality importance, leading to unfair conclusions. To delve into it further, let us consider a feature vector $\mathcal{F}(\mathcal{V}^n) = \{f_1^n, f_2^n, \ldots, f_m^n\}$, where $\mathcal{F}(\mathcal{V}^n)$ represents the feature vector of the $n$-th node and $f_i^n$ represents features derived from the $i$-th modality. To gauge the effect of eliminating the $i$-th modality, we create a new feature vector $\mathcal{F}(\mathcal{V}_{-i}^n) = \{f_1^n, f_2^n, \ldots, f_{i-1}^n, f_{i+1}^n, \ldots, f_m^n\}$ which includes information from all modalities except the $i$-th one. After creating this new feature vector for all nodes, the similarity between nodes will change. This means that nodes $n$ and $k$ that were initially similar may no longer be similar, and vice versa:

$$\begin{cases} Sim\left(\mathcal{F}(\mathcal{V}^n),\ \mathcal{F}(\mathcal{V}^k)\right) = \alpha < \tau \Rightarrow \mathcal{V}^n, \mathcal{V}^k \text{ neighbors} \\ Sim\left(\mathcal{F}(\mathcal{V}^n_{-i}),\ \mathcal{F}(\mathcal{V}^k_{-i})\right) = \alpha > \tau \Rightarrow \mathcal{V}^n, \mathcal{V}^k \text{ not neighbors} \end{cases} \qquad (1)$$

Here, $Sim(\cdot)$ represents the similarity function between two vectors, and $\tau$ represents the threshold below which nodes with similarity values are considered neighbors. Researchers face two choices in this context: (i) they can reconstruct the graph structure based on the new similarities, or (ii) they can retain the original graph structure and continue with the analysis. In the first scenario, where the graph structure is reconstructed, comparing the performance of the graph-based model on this new graph with the original one would not provide a fair assessment of the eliminated modality's impact. However, in the second scenario, maintaining the structure of the initial graph poses a risk of connecting dissimilar nodes. This could confuse the model by incorporating information from irrelevant neighbors, potentially leading the model away from its optimal performance. Therefore, removing modalities in graph-based studies, particularly when edges are determined by node features, is not a fair method for assessing the impact of each modality.

## 5.2. Category II: Non-Graph XAI

In addition to evaluating the importance of each modality, several studies have focused on determining the significance of various components of their constructed graph. This includes assessing the importance of an individual node, the edges between nodes, and features derived from different modalities for each node. To achieve this, researchers have employed several post-hoc XAI techniques originally developed for general machine learning, rather than specifically for graph-based models. For example, the saliency map technique [46, 47], which relies on gradient calculation, was used by several studies [22, 24, 25] to assess the importance of each node's feature. Bi et al. [25] also used this technique to determine the importance of graph nodes, where each node represents a specific brain ROI. Another gradient-based XAI technique, which integrated gradient analysis [48], was used by Huo et al. [32] to unveil the importance of each genomic feature employed in their study. Additionally, Qu et al. used Grad-RAM [20], another gradient-based technique, to identify the importance of each node in the graph, where each node also represents a specific brain ROI. Qu et al. [20] further provided edge-level interpretability using an edge-masking technique, which involved systematically removing one edge at a time from all patient graphs and retraining the model to observe the resulting performance changes, thereby determining the significance of each edge. Zhang et al. [28] and Pfiefer et al. [30] used SHAP [45], a technique based on Shapely values, to illustrate the importance of each feature and nodes, respectively. Sensitivity analysis [49], a common technique to determine feature importance in neural networks, by multiplying network weights and feature standard deviation, was used by Li et al. [38] to identify feature importance in their study. **Figure 5** provides detailed visualizations of each of these techniques. **Table 2** summarizes the various non-graph XAI methods along with their corresponding formulae employed in different studies.

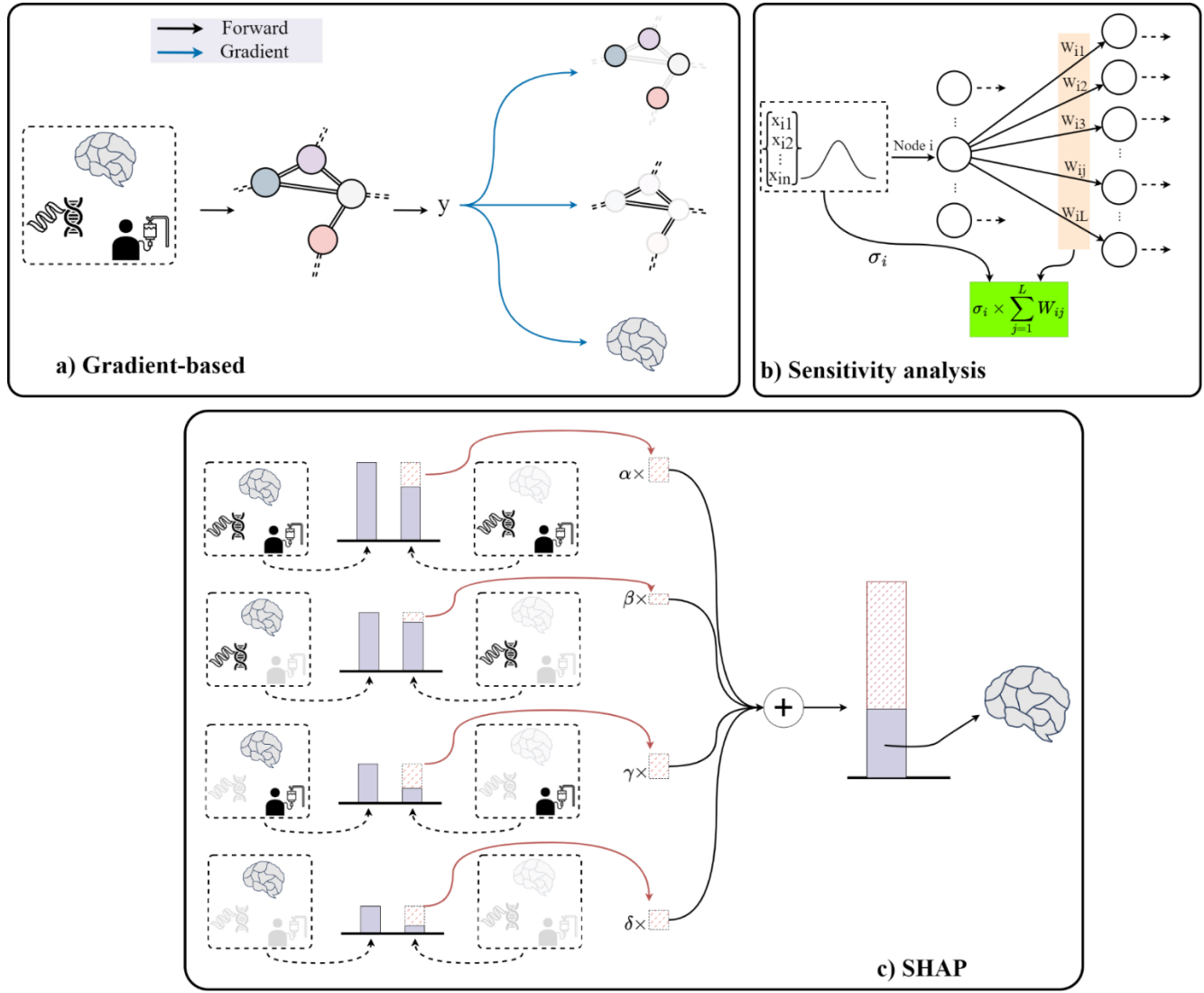

**Figure 5**. **An overview of various non-graph XAI techniques used in graph-based studies to gain insights from the model. a)** Gradient-Based Method: This technique is used to assess the importance of different components in a graph-based model. It involves calculating the gradient of the model's predicted value with respect to various elements, such as graph nodes, edges, input features, or modalities. By analyzing these gradients, researchers can identify which components most significantly influence the model's predictions. **b)** Sensitivity analysis: This is a common method for determining the importance of features in a neural network. It involves calculating the standard deviation of a specific feature across all samples and multiplying it by the sum of the weights associated with that feature in the trained network. This product reflects the feature's overall importance to the model's performance. **c)** SHAP (Shapley

Additive Explanations): SHAP is a widely used method for enhancing model interpretability. It calculates the model's performance with and without specific modalities (e.g., brain ROIs on the left side). The difference in performance when a modality is included versus when it is excluded is computed across all possible modality combinations. These differences are then combined linearly to quantify the contribution of each modality to the model's overall performance.

**Table 2.** The summary of various non-graph XAI techniques employed in the surveyed studies alongside their respective formulas.

| XAI Technique | Explanation | Formula | Denotation |
|---|---|---|---|
| Gradient saliency map | A method to determine the importance of each entity in a model by calculating the gradient of the model's prediction with respect to that specific entity. | $\frac{\partial \hat{y}}{\partial E}$ | $\hat{y}$: The score vector produced by the model., $E$: The specific value associated with an entity in the model, which could refer to a node, its features, or edges. |
| Integrated gradient [48] | A refined version of the gradient saliency map, mitigates sensitivity breakdown commonly observed in basic gradient saliency maps. | $(x_i - \acute{x}_\iota) \int_{\alpha=0}^{1} \frac{\partial F(\acute{x} + \alpha(x - \acute{x}))}{\partial x_i} d\alpha$ | $x$: real input, $\acute{x}$: baseline (Black image for imagery, zero embedding vector for text), $F$: model's function |
| Grad-RAM [20] | A method to track the gradient of the model with respect to the prediction value | $\frac{1}{NC} ReLU \left( \sum_{n=1}^{N} \sum_{C=1}^{C} \frac{\partial \hat{y}_n}{\partial E} E \right)$ | $N$: Number of samples, $C$: number of feature channels, $\hat{y}_n$: the label for the $n$th subject, $E$: The specific value associated with an entity in the model, which could refer to a node, its features, or edges. |
| SHAP [45] | A method that integrates multiple additive XAI techniques, offering feature importance explanations that are more intuitive for humans. | $\sum_{\acute{z} \subseteq \acute{x}} \frac{|\acute{z}|! \ (M - |\acute{z}| - 1)!}{M!} [f_x(\acute{z}) - f_x(\acute{z} \backslash i)]$ | $|\acute{z}|$: the number of non-zero entries in $\acute{z}$, $\acute{x}$: simplified input corresponds to the original input $x$, $M$: the number of simplified input features, $f_x$: model's function, $\backslash i$: excluding feature $i$ values from the input. |
| Sensitivity analysis [49] | A method employed to quantitatively describe the significance of | $\sigma_i \times \sum_{j=1}^{L} |W_{ij}|$ | $\sigma_i$: standard deviation of a feature $x_i$, $W_{ij}$: the connection weight of the input nodes to the |

| input variables in neural networks. | output nodes, $L$: the number of nodes in the next layer |

## 5.3. Category III: Graph-Based XAI

While post-hoc non-graph XAI techniques can offer valuable insights into the importance of various components in graph-based studies, it is important to recognize their limitations. Gradient-based XAI techniques, for example, have been shown to be misleading in certain scenarios [50] and can suffer from issues like gradient saturation [48, 51]. Additionally, these techniques are not specifically designed for graphs where inputs are often discrete, which further limits their effectiveness when applied to graph-based models [52]. To overcome these challenges, researchers have employed XAI techniques specifically tailored for graph-based models.

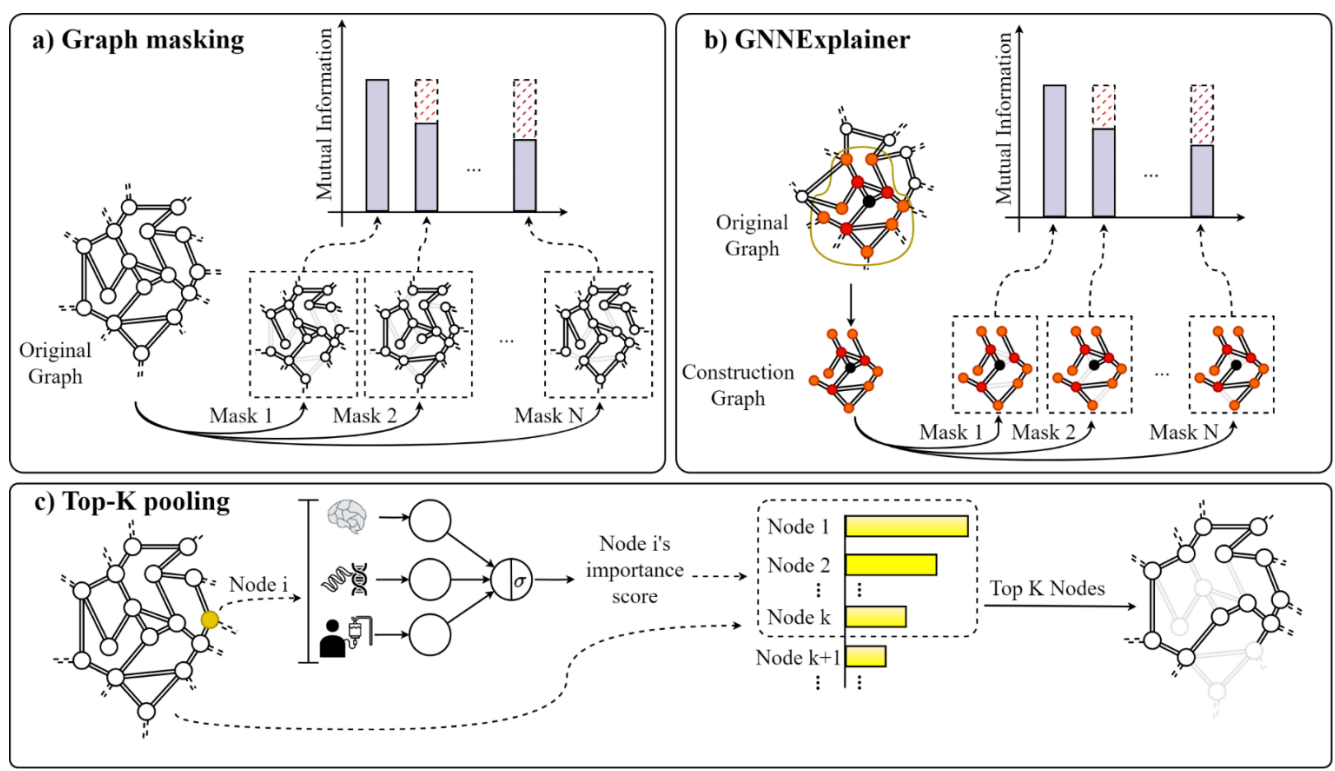

**Figure 6. An overview of graph-based XAI techniques used to gain insights from graph-based models**. **a)** Graph masking: This technique involves applying a mask (either an edge mask or a feature mask) to the original graph structure. The mutual information between the masked graph and the true label distribution is then calculated. The masked graph with the highest mutual information value is identified as the most important or informative graph. **b)** GNNExplainer: Similar to the graph masking method, GNNExplainer applies masks to the graph to uncover important components, such as nodes, edges, or features, that significantly influence the model's predictions. **c)** Top-k Pooling: In this method, each node's features are passed through a single-layer multi-layer perceptron (MLP) to calculate the importance of each node. The nodes are then ranked based on their importance scores, and the top-k nodes are selected to construct the final graph.

Zhou et al. [23] employed a general graph masking technique to identify the most significant subgraph and subset of features within their model. They specifically developed optimal masks for both the graph structure and the feature vector, allowing them to distinguish the most important subgraph and feature subset. Additionally, GNNExplainer [52], another masking-based technique, was employed by Pfeifer et al. [29] and Tang et al. [44] to pinpoint the most crucial nodes and features within a graph during the prediction process for a specific sample. The primary difference between the general graph masking technique and GNNExplainer lies in their scripts. While GNNExplainer focuses on the construction graph of a particular node—essentially a subset of the original graph that directly influences the output for that node—the general graph masking technique considers the entire original graph. GNNExplainer offers computational advantages, especially in practical applications involving large graphs with potentially millions of nodes. Another technique used to determine the importance of each node in an input graph is the top-k pooling technique [53]. In this method, the importance score of graph nodes is computed using multi-layer perceptron (MLP) models. This approach, which offers node-level interpretability, was applied by Sebenius et al. [21] to brain ROIs to identify the most important ROIs for diagnosing patients with schizophrenia. The details of the graph-based XAI techniques are also demonstrated in **Figure 6**, while **Table 3** summarizes different graph-based XAI techniques along with their corresponding formulae employed in the surveyed studies.

**Table 3.** Summary of various graph-based XAI techniques employed in the surveyed studies alongside their respective formulas.

| XAI Technique | Explanation | Formula | Denotation |
|---|---|---|---|
| Graph masking | A method systematically eliminates certain parts of the graph structure along with their corresponding feature vectors to identify the most crucial ones | $G_s = A \odot P_A, X_s = X \odot P_X$ | $G_s$: Target important subgraph of the original graph, $X_s$: Target important subset of feature vector $X$, $A$: Adjacency matrix, $P_A$: Learnable edge importance probability, $P_X$: Learnable feature importance probability |
| GNNExplainer [52] | A method similar to graph masking technique except it focuses only on the construction graph to find the most important subgraph instead of analyzing the whole graph | $-\sum_{c=1}^{C} \mathbb{1}[y = c] \log P_\phi(Y = y \mid G$ $= A_c \odot \sigma(M), X = X_s \odot F)$ | $c$: a specific class among a total number of $C$ classes, $A_c$: Computation graph's adjacency matrix, $\sigma$: the sigmoid function, $M$: Learnable graph mask, $F$: Learnable feature mask |

| Top-k pooling [53] | A method to calculate an importance score for each node in a graph | $\delta((\|p\|)^{-1} X p)$ | $\delta$: a non-linear function, $p$: a learnable vector, $X$: input samples, $\|\cdot\|$: $L_2\ norm$ |
|---|---|---|---|

## 5.4. Category IV: Inherent Interpretability

While post-hoc XAI techniques are valuable for gaining insights into the components of complex models, interpretable models offer inherent transparency by allowing tracking of the data processing steps. This transparency enables understanding of how different components collaborate to generate predictions, clearly showing how features, nodes, and edges contribute to the model's outputs. In the realm of multimodal medical data integration using graph-based models, several studies have developed Inherent interpretable models. The elements used in these studies can be classified into the following three categories.

**Attention Mechanism:** Several studies integrated attention mechanisms, which assign higher attention weights to more important components. In the study conducted by Yang et al. [19], graph inputs are first processed through a multi-head attention layer to compute attention weights for each node. By employing gradient sensitivity analysis to these weights, the study reveals specific features and their interactions pertinent to the classification process. In another study, Ouyang et al. [43] introduced an attention layer within their model to elucidate the significance of each modality. They also incorporated inner product regularization on the feature indicator matrix to emphasize the more critical features within each modality. Similarly, Huo et al. [32] also employed attention mechanisms to identify significant clinical and pathological features within the corresponding graphs in their study. In another work, Ma et al. [33] utilized the attention mechanism in their model to determine the importance of genes within specific cells. Kazi et al. [40] developed IAM, an attention-based component in their model designed to reveal the importance of each input feature. Additionally, the authors introduced a novel loss function tailored specifically for this component, enhancing the IAM's ability to detect more important features. Lastly, in the model proposed by Bintsi et al. [39], features are initially processed through an attention layer. Subsequently, the weighted features, constructed by multiplying feature values, and the weighted features—created by multiplying feature values by their corresponding attention weights—serve as the feature vector for each node (patient). Once the model is fully trained, the attention weights reveal the importance of each feature, identifying the most crucial ones.

**Graph Attention:** Beyond the general attention layers used in the above studies, some researchers incorporated graph-specific attention mechanisms to enhance the interpretability of their model. In the study by Keicher et al. [35], they used a GAT [17], where the attention weights shed light on how neighboring nodes (patients) influence the classification of a specific node (patient). This approach is particularly valuable in pandemic scenarios, such as the COVID-19 study they conducted, offering valuable insights into population interactions and potential outcomes for individual patients. Similarly, Safai et al. [24] also employed GAT to understand the significance of nodes (brain ROIs) in the model's prediction, aiming to elucidate the interactions between various brain regions. Additionally, Li et al. [34] used GAT to identify the importance of different nodes (drug-protein pairs) in their study. Xiao et al. [42] leveraged GAT in their proposed model to determine each node's (patient) contribution to the prediction of outcomes for other patients. Additionally, Kaczmarek et al. [18] used the graph transformer, another graph-specific attention mechanism, to analyze the significance of interactions between nodes (specifically mRNA-miRNA interactions) in detecting various types of cancers. They further calculated the average edge weights associated with each node to identify key nodes (mRNA or miRNA) relevant to specific cancer classification tasks.

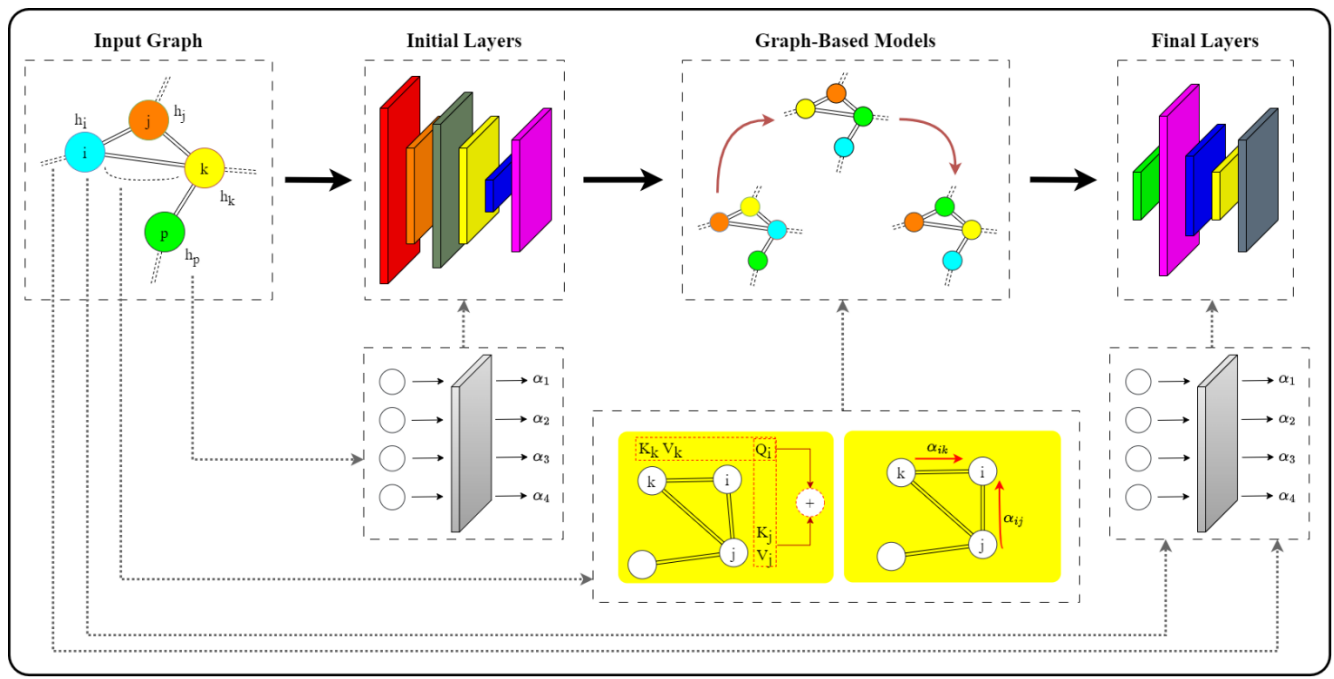

**Figure 7. A general guideline for developing an interpretable graph-based platform on multimodal medical data.** It is important to note that the purpose of interpretability and the study's objective significantly influence the type of elements to use and their placement within the model.

**Other Methods:** Some other researchers have developed interpretable graph networks without relying on attention mechanisms. Kan et al. [26] designed a neural network centered on the nodes and edges within their graph, determining the importance of each feature and node by analyzing their associated weights in the network. Similarly, Bi et al. [41] leveraged the weights of graph convolutional filters as indicators of feature importance in their model. Lastly, Pfiefer et al. [30] incorporated decision trees into their model and capitalized on their inherent interpretability to unveil the significance of each modality and feature in their research.

Depending on the specific objective of a study, one or a combination of the aforementioned approaches can be useful in developing interpretable graph-based models. To guide future research on interpretable graph-based models for medical multimodal data, we provide a general guideline depicted in **Figure 7**. As depicted, the study's objective determines which elements to use and where to place them within the model. If the goal is to identify the most important nodes and edges, using attention mechanisms in the final layers is recommended. Conversely, if the aim is to understand the importance of features, employing an attention mechanism in the initial layers is recommended. Additionally, using GAT and GTN as graph-based models can greatly aid researchers in understanding node interactions within a graph and determining which nodes have the most impact on a specific node's prediction.

Indeed, as demonstrated in the aforementioned studies, interpretable models often use specific layers to highlight the importance of each component within the model. However, to maintain transparency in data processing and insights, these models are typically kept relatively simple. This creates a trade-off between enhancing model performance and ensuring transparency, which requires careful consideration during the model design phase.

Overall, the majority of studies implementing graph-based models on multimodal medical data have relied on conventional XAI techniques such as gradient-based methods, SHAP, and other approaches originally designed for general machine learning rather than specifically for graphs. A smaller subset of studies has employed techniques like GNNExplainer, which are tailored for graph-based model explainability. Additionally, several studies have developed interpretable models, many of which incorporate attention mechanisms as a key component. In the next section, we present a benchmark analysis that compares these methods in practice, highlighting their advantages and limitations.

### 5.5. Benchmark Analysis

We conducted a benchmarking analysis to evaluate the performance and practicality of the previously described XAI techniques when applied to a graph-based model. Specifically, we focused on MOGONet, a graph-based model developed by Wang et al. [36] for biomedical classification tasks across various diseases. In our study, we replicated the MOGONet architecture using the ROSMAP dataset [54], which is curated for Alzheimer's disease (AD) classification. The dataset comprises features from three distinct modalities—mRNA expression data, DNA methylation data, and microRNA expression data—each processed through its respective graph-based model within the MOGONet framework. Final classification outcomes were obtained by integrating the predictions generated by these individual models.

To assess feature importance, the original study in MOGONet employed a modality elimination technique (Category I), wherein the performance impact of excluding specific features was measured. Building on this approach, we extended the analysis by implementing additional XAI techniques from Categories II and III to gain deeper insights into the relative significance of the dataset's features. Category IV methods were excluded from this analysis, as these inherently interpretable models are not model-agnostic and thus cannot be retrofitted onto pre-existing architectures like MOGONet. For this benchmarking analysis, we selected gradient saliency Map, SHAP, and sensitivity analysis from Category II. From Category III, we incorporated graph masking, a generalization of GNNExplainer, into the analysis due to its broader applicability in identifying important features and subgraphs. However, the top-k pooling method, which is specifically designed to assess node importance, was excluded as it is not directly applicable to feature-focused analyses.

After training, we used the trained model's parameters and the feature distributions from the test set to calculate the importance of each feature using sensitivity analysis. This technique was relatively fast because it didn't require backpropagation or additional model training. To compute feature importance using gradient saliency maps, we calculated the gradient of the model's output for each test sample with respect to each feature. The feature importance was revealed by averaging these gradients across all test samples. For the graph masking technique, we aimed to find a mask that, when applied to the graph, maximized mutual information with the non-masked graph. We started with a randomly initialized mask and optimized it over 100 epochs using the binary cross-entropy cost function. This method was more time-consuming than the previous two, as we needed to find the optimal mask for each test sample individually. Finally, we used Kernel SHAP to compute feature importance. We generated 100 weighted neighbors for each test sample, with the weights calculated using the equations proposed by Lundberg et al. [45]. By training a logistic regression model on these generated samples, we determined the feature importance for each sample. For global interpretability, we averaged these importance values across all patients.

We first compared the top genes identified by each XAI method. As **Figure 8a** showed all four methods—SHAP, sensitivity analysis, gradient saliency, and graph masking—identify partially overlapping yet distinct sets of top genes, reflecting the inherent variability in how each explainability algorithm quantifies feature importance. SHAP and sensitivity analysis share more genes in common (e.g., *CTB-171A8.1*, *RP11-552D4.1*, *OR7A5*), consistent with both methods having a relatively broad coverage of biologically relevant signals. Gradient saliency highlights genes such as *NPNT*, *SYTL1*, and *ANKRD30B*, which do not appear prominently in the other lists, suggesting its sensitivity to different aspects of model gradients. Graph masking surfaces genes associated with membrane transport and cell signaling (*SLC25A18*, *GPIHBP1*, *CLCA4*, *SCD*), reflecting its unique strategy of learning a mask that pinpoints the most relevant subsets of the graph structure.

To assess how each interpretability method's top 30 genes compared against random chance, we calculated the average importance score of those top 30 genes and then generated a null distribution by repeatedly (1,000 times) sampling 30 genes at random (**Figure 8b**). In the figure, the ratio indicates how far this observed average lies from the mean of the random distribution. Notably, gradient saliency shows the largest offset, implying that its top-ranked genes are especially unlikely to have been identified by chance, whereas sensitivity analysis yields a smaller offset but still remains distinctly above the random null distribution. Graph masking and SHAP occupy intermediate positions, further underscoring that each algorithm captures important gene sets in different ways.

We then conducted gene ontology (GO) analysis on the top genes identified by each XAI technique and extracted the corresponding GO terms, keeping only those with p-values ≤ 0.01. **Figure 8c** presents the GO terms alongside the XAI techniques that identified them. As the figure shows, SHAP identified ten out of eighteen Alzheimer-related GO terms (unique GO terms identified by all methods), while sensitivity analysis identified six. In comparison, graph masking and gradient saliency each detected only two Alzheimer-related GO terms. Interestingly, each XAI method highlighted distinct yet sometimes overlapping GO terms relevant to AD. SHAP detected multiple AD-associated processes, including *positive regulation of cytosolic calcium ion concentration*, *endocrine system development*, and *temperature homeostasis*, each tied to neuronal dysfunction and metabolic dysregulation in AD. Sensitivity analysis captured partially overlapping GO terms (e.g., *endocrine system development*) while adding *ensheathment of neurons*, *gliogenesis*, and *forebrain development*, reflecting broader coverage of neurodevelopmental processes. Gradient saliency identified fewer terms, but still pinpointed *positive regulation of cytosolic calcium ion concentration* and *carbohydrate transport*, both linked to neuronal health and metabolic deficits in AD. Graph masking introduced unique findings like *14-3-3 protein binding*, relevant to tau pathology, and *actin filament*, associated with cytoskeletal disruptions.

We then repeated the same analysis on the 15 unique pathway terms identified by all the methods (we only considered those pathways with p-values ≤ 0.01). Notably, none of these retained pathways were identified by gradient saliency, seven identified by SHAP, while both sensitivity analysis and graph masking methods identified six pathways (**Figure 8d**). SHAP and sensitivity analysis both prominently identified pathways involving G protein-coupled receptor (GPCR) signaling (e.g., GPCR ligand binding, GPCR downstream signalling, Class A/1 Rhodopsin-like receptors, signaling by GPCR)—processes closely linked to synaptic transmission and neurodegeneration in AD. Sensitivity analysis, in particular, highlighted dopaminergic neurogenesis and hematopoietic stem cell differentiation, reflecting broader aspects of neuronal and immune system development relevant to AD. SHAP, meanwhile, singled out the cAMP signaling pathway, another key mediator of neuronal plasticity and survival that has been implicated in AD pathophysiology. In contrast, graph masking emphasized more general transport pathways (e.g., SLC-mediated transmembrane transport, bile secretion, gastric acid secretion) and metabolic processes, which may indirectly relate to AD through systemic metabolic dysregulation but are less directly established in AD literature compared to GPCR signaling or dopaminergic function.

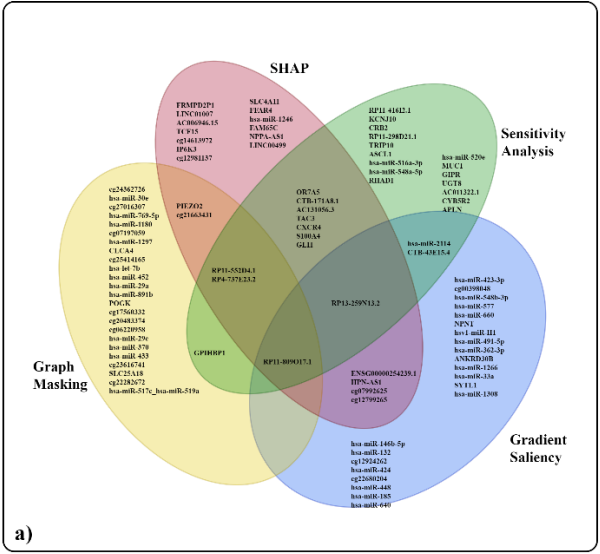
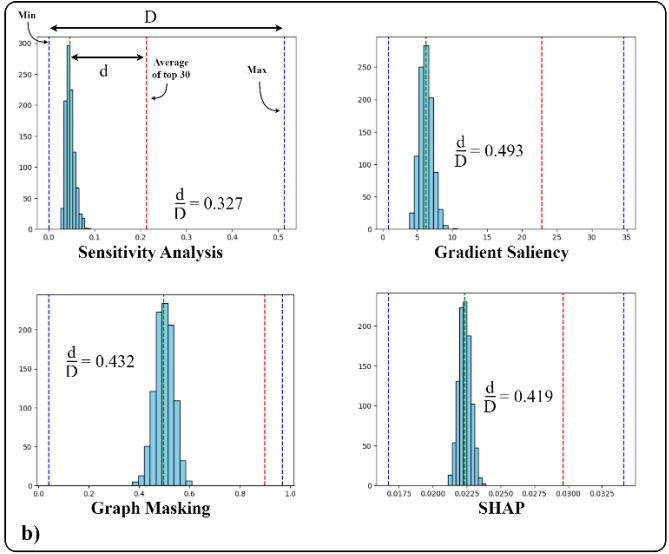
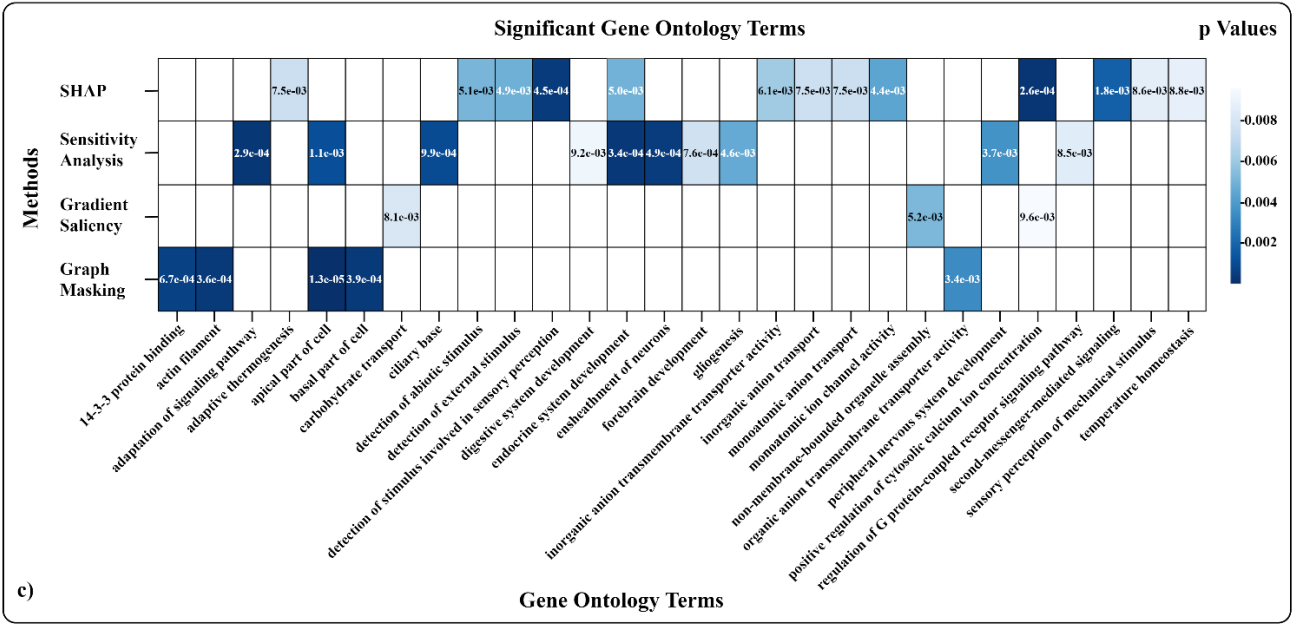
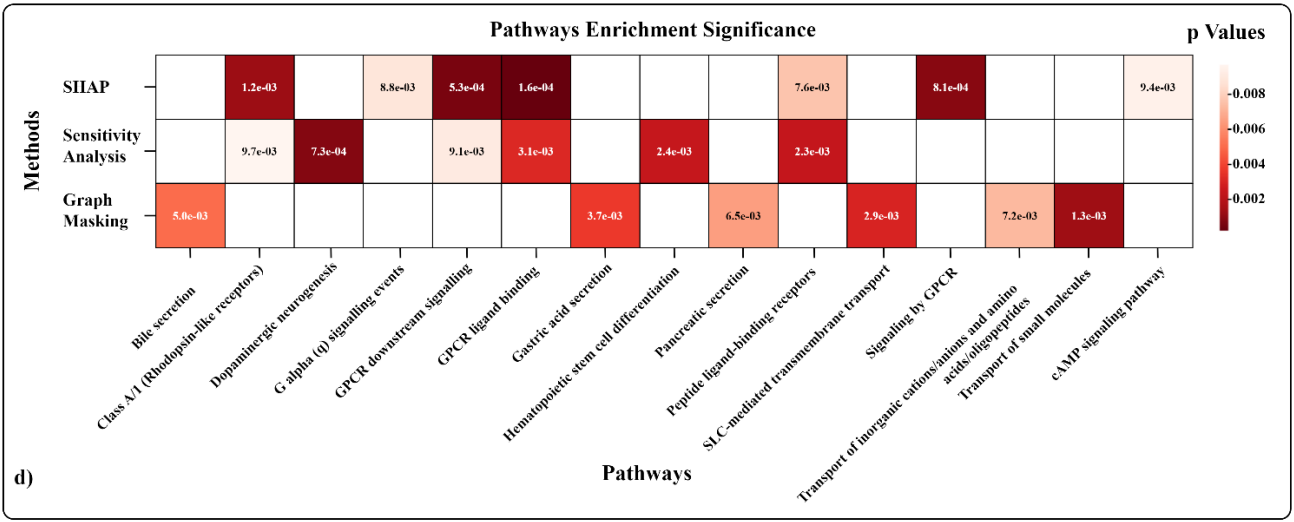

**Figure 8. Benchmarking analysis**. **a)** Top 30 features captured by different XAI techniques from category II and III considered as the most important features for Alzheimer disease classification. As shown in the figure, SHAP and gradient saliency were roughly able to detect a combination of the most important genes found by other techniques, as the most important genes for Alzheimer disease classification. **b)** A permutation analysis was performed on the important features identified by each technique. Specifically, 30 features were randomly chosen from the entire feature set, and their average importance scores were calculated. This random selection process was repeated 1000 times, creating a distribution of average importance scores, which was visualized in a histogram. The average importance score of the top 30 ranked features (shown by a red dotted line) was then compared to the mean of the permutation-based score distribution to assess the significance of these selected features. The results show that the gradient saliency technique outperformed the other techniques. **c)** In the GO analysis, we identified several GO terms, retaining only those with a p-value ≤ 0.01. These terms, along with their p-values and the XAI techniques that identified them, are shown. **d)** Pathway analysis was also conducted, keeping only pathway terms with a p-value ≤ 0.01 and an overlap > 1; notably, none of these terms were identified by gradient saliency. The retained pathway terms, the XAI techniques that found them, and their p-values are displayed**.** XAI: eXplainable Artificial Intelligence, SHAP: SHapley Additive exPlanations GO: Gene Ontology

## 6. Discussion

In this paper, we technically examined how graph-based models can facilitate multimodal medical data integration and subsequently improve interpretability in tasks such as disease classification. By benchmarking various explainability approaches—ranging from post-hoc XAI to inherently interpretable models—we demonstrated distinct trade-offs between computational overhead and the granularity of explanations. Our findings indicate that, while gradient-based and sensitivity-based methods can efficiently spotlight important features, more sophisticated approaches like SHAP or graph masking provide deeper interpretative insights, albeit at higher computational cost.

Graph-based models are gaining significant interest from researchers in integrated medical studies. While many studies have focused on either node-level or graph-level analysis, there remains potential to explore graph-in-graph techniques, which could unlock the full power of graph-based models [1]. In these studies, each sample in a dataset is represented as a node within a larger graph, and each individual sample (node) is itself a graph, allowing for simultaneous node-level and graph-level analysis. This approach maximizes the capabilities of graph-based models in practice.

A hypothetical application is to model an entire patient cohort as a large graph, where each patient is a node, while each patient's internal molecular or imaging data forms its own subgraph. For instance, in cancer analysis, every patient (node) can include a subgraph of gene–gene interactions or imaging-based ROI connectivity. This hierarchical design enables global population-level insights (e.g., grouping patients with similar molecular pathways) while capturing local, detailed interactions within each patient's subgraph. Such a two-level representation can potentially reveal key markers of disease progression or response to therapy that might remain hidden in single-level graph analyses.

In practice, time and computational resource limitations make it crucial to carefully choose the appropriate post-hoc XAI technique. For example, using SHAP based on Shapley values can be both computationally expensive and time-consuming (in our benchmarking analysis with 600 features, generating the final report could take several weeks). Given these challenges, inherently interpretable models are becoming more appealing. **Figure 7** provides a step-by-step guide for developing transparent graph-based models suitable for multimodal medical data. However, it is important to note that interpretable models are often less accurate and may perform sub-optimally. Therefore, researchers must weigh their study's objectives, the available computational and time resources, and the required level of performance when deciding which technique or combination of techniques to implement [55]. Evidence from similar fields [56-58] suggests that this performance–interpretability trade-off can sometimes be reduced via hybrid approaches (e.g., combining attention-based layers with simpler interpretable modules or augmenting inherently interpretable models with lightweight post-hoc explanations). In contexts where patient outcomes hinge on actionable insights, prioritizing a slightly lower predictive accuracy but higher interpretability may be clinically justified [59].

While interpretable graph-based models applied to multimodal medical data have found broad utility in cancer analysis [27, 29, 31, 32], existing studies have predominantly focused on a limited subset of cancers such as breast [28, 38] and kidney [30]. This highlights an open area of research for extending these techniques to analyze other prevalent cancers like prostate, skin, and thyroid cancers. Similarly, studies leveraging interpretable graph-based models for disease diagnosis have primarily concentrated on a narrow group of diseases including COVID-19 [35], Alzheimer's [23, 25, 40, 41], and Parkinson's [24, 37]. It is advisable to explore the application of these models on multimodal medical data for detecting other widespread diseases such as malaria, AIDS, and hepatitis. Furthermore, while most studies have developed models for classification or regression tasks, there is a need to expand this field of study to include image segmentation and object detection in medical imagery data to assess their potential efficacy. Such diversification could not only confirm the generalizability of graph-based explainable approaches but also expose task-specific challenges—e.g., the different ways imaging data, lab tests, and patient demographics interact in complex diseases—thus spurring methodological innovations. When applying these methods to underexplored domains, the availability and integration of multimodal datasets can pose additional challenges. For instance, large-scale imaging data for diseases like malaria or AIDS may require specialized preprocessing pipelines, whereas clinical data for hepatitis might demand domain-specific approaches to handle heterogeneity in liver enzyme tests, patient demographics, and treatment histories. Graph-based architectures that can flexibly fuse multiple data types—such as sequences, images, and clinical variables—stand to offer considerable value in such scenarios.

In the graph construction step of existing studies, some have employed background knowledge [27-31] to establish edges between nodes, while others have relied on similarities between node features calculated using functions such as cosine similarity [42, 43], Euclidean distance [38, 40, 41, 44], and Pearson correlation [20, 22, 25, 37]. There is an open area of research to explore constructing graphs based on knowledge graphs containing multimodal biomedical data. Additionally, researchers could investigate the use of alternative similarity functions like Jaccard similarity, Dice similarity, or a combination of knowledge graphs and these similarity functions. Beyond static graph construction, future directions could also involve dynamic approaches where knowledge-graph edges are updated during training or are conditioned on specific contexts—e.g., disease subtype or patient demographic. This iterative refinement might identify latent relationships that remain unseen in purely static or correlation-based graphs.

While some studies have employed post-hoc explainable methods tailored for graph networks, such as GNNExplainer [52], there remains potential for future studies to explore alternative explainable techniques designed specifically for graph networks [60-62]. In future studies, employing such techniques may prove more suitable for diverse constructed graphs, potentially leading to enhanced interpretability of the models at a higher level. Table 4 offers a comprehensive overview of all the studies examined in this paper, providing detailed information on each.

Our benchmarking results also reveal that each XAI method offers unique strengths and perspectives in uncovering AD-related biology. SHAP and sensitivity analysis consistently identified more established AD-associated GO terms and pathways—particularly those involving GPCR signaling, neurogenesis, and protein binding—making them especially useful for studies requiring high interpretive depth and alignment with known disease mechanisms. Gradient saliency and graph masking, while faster or more novel in approach, each contributed distinct insights (e.g., *14-3-3 protein binding* and metabolic transport pathways) that highlight lesser-studied facets of AD pathophysiology. Permutation tests further confirmed that all four methods generated top-ranked genes well above random expectations, with gradient saliency producing notably distinct importance scores. In terms of computational cost, SHAP and graph masking required more processing time, whereas gradient saliency and sensitivity analysis ran faster but sometimes yielded fewer established AD indicators. Overall, these findings underscore the complementary nature of the methods: combining multiple XAI techniques can provide a more comprehensive understanding of the complex processes driving Alzheimer's pathology while balancing the trade-offs between interpretive depth and computational demand.

Moreover, emerging XAI methods—such as counterfactual reasoning [63] and contrastive explanations [64, 65]—hold particular promise for revealing minimal changes to node features or edges that alter model outputs, thus guiding personalized interventions. By illuminating the precise pathways through which data influence outcomes, these methods can amplify the clinical impact of graph-based analyses in settings where pinpointing the root cause of a diagnostic decision is paramount.

## 7. Future directions: LLM-Integrated graph models

While interpretable graph-based models have significantly advanced the analysis of multimodal biomedical data, integrating Large Language Models (LLMs) such as GPT-based architectures offers a promising path toward more thorough and user-friendly solutions. LLMs excel at interpreting unstructured text—including clinical notes [66], scientific publications [67], and patient records [68]—and can transform these insights into structured knowledge that bolsters both graph construction and interpretability. A key advantage lies in the ability of LLMs to identify entities (e.g., genes, proteins, symptoms) and the relationships between them, subsequently mapping this information onto nodes and edges in a knowledge graph [69]. By iteratively updating these relationships during training, one can build more dynamic graph representations that capture context-specific connections beyond traditional similarity metrics (e.g., Pearson correlation).

Another dimension where LLMs can improve interpretability is in generating natural language explanations for model outputs. Graph-specific XAI methods such as GNNExplainer [52] or graph masking typically yield numeric feature attributions or visual node connectivity, but they rarely articulate why a subgraph might be biologically relevant. An LLM, prompted with queries about the significance of specific subgraphs or node clusters, can consolidate domain knowledge from research abstracts or curated databases to provide textual rationales. These can include potential disease associations, known molecular pathways, or drug–gene interactions, thereby humanizing the explanation process and making it more accessible to clinical stakeholders.

LLMs can also produce semantic embeddings that capture extensive contextual information from large-scale biomedical texts [70]. These embeddings, aligned with numeric features like gene expression levels or imaging metrics, allow a hybrid model to emphasize both the quantitative importance of certain edges and their qualitative significance as gleaned from domain literature. For example, two genes may be statistically correlated in expression data, but the LLM-derived context might reveal that they are also co-mentioned in numerous studies on AD, thus reinforcing their importance within a GNN's message-passing operations.

A compelling application of LLMs and GNNs is interactive querying. Clinicians or researchers could ask natural language questions—for instance, "Which molecular markers are most predictive of advanced Parkinson's disease?"—and receive evidence-based responses that reference both the underlying subgraph structure and the relevant biomedical literature. This approach can expedite hypothesis generation, guide experimental design, and bridge the gap between domain experts and data-driven insights.

Despite these potentials, several technical challenges accompany LLM-integrated graph models. Training or fine-tuning large LLMs can be resource-intensive, requiring specialized hardware and careful optimization strategies [71]. There is also the risk of "hallucination," where an LLM invents facts unsupported by scientific evidence [72]. This issue necessitates robust validation pipelines, perhaps involving cross-referencing with established knowledge bases or subject-matter experts. Additionally, privacy concerns arise when handling personal health records and clinical data, warranting advanced anonymization and security measures, including privacy-preserving data linkage. Bias in pretrained text corpora further complicates clinical fairness, as it can inadvertently skew predictions or explanations [73].

## 8. Conclusion

This study offers a comprehensive overview of research on interpretable graph-based models applied to multimodal biomedical data. Integrating data from various modalities has proved essential for thorough analyses and improved medical outcomes, and graph-based models—through their ability to capture spatial relationships and leverage domain knowledge—have emerged as robust solutions in this space. Yet, the interpretability of these models remains critical, particularly in sensitive healthcare contexts where trust, accountability, and ethical considerations are paramount. Over the past few years, there has been a marked shift toward employing interpretable methods, either through inherently interpretable architectures or post-hoc explainability techniques that elucidate model outputs. As the only review focusing specifically on interpretable graph-based models for multimodal biomedical data, this paper takes a pivotal step in mapping existing methodologies and providing a reference point for both seasoned researchers and newcomers in the field. Beyond summarizing state-of-the-art approaches, our work highlights uncharted directions such as graph-in-graph architectures, knowledge-graph-driven edge construction, and the exploration of novel XAI methods dedicated to graph neural networks. Moreover, the integration of LLMs opens promising new frontiers in augmenting both data representation and interpretability, allowing us to enrich graph structures with information extracted from unstructured text, generate semantically rich node embeddings, and present human-readable rationales. Going forward, researchers can capitalize on these emergent techniques to further expand the range of diseases and tasks tackled by interpretable graph-based models, stretching beyond the current focus on a few cancer types and neurodegenerative diseases. By combining powerful LLM-based tools with advanced GNN frameworks, future models can yield clearer insights while addressing the practical constraints of time, computational resources, data biases, and privacy. Overall, this review—together with the directions outlined herein—lays a comprehensive foundation for a new era of explainable, data-driven solutions in healthcare, poised to substantially accelerate medical research and clinical decision support.

### Method and materials

### Gene ontology and pathway analyses

We employed Over-representation Analysis (ORA) via WebGestalt (https://2019.webgestalt.org/) [74] to investigate gene ontologies and biological pathways associated with protein-coding genes. This approach aimed to identify fundamental biological processes, molecular functions, and cellular components. For GO terms, we consulted non-redundant databases encompassing the three major domains—biological processes, cellular components, and molecular functions. We additionally included pathway annotations from KEGG, Reactome, Panther, and WikiPathway to capture a broad spectrum of potential molecular mechanisms. All GO terms and pathways were prioritized based on their False

Discovery Rate (FDR) values (p < 0.01), thereby minimizing the likelihood of false positives and highlighting the most statistically significant findings.

**Benchmarking XAI approaches on MOGONet**

To replicate MOGONet on the ROSMAP dataset, we strictly followed the methodology implemented by its authors. The ROSMAP dataset comprises 351 samples, with 169 (48%) classified as normal cases and 182 (52%) as AD cases. The dataset is divided into training and testing sets, with 70% (245 samples) used for training and 30% (106 samples) used for testing. The division is stratified to ensure that Alzheimer's cases consistently represent 52% of both subsets. Each sample includes three modalities—DNA methylation, mRNA expression, and microRNA expression—each represented by 200 features, which serve as input to the model. Each modality is processed by its respective graph-based model. The initial predictions from these modality-specific graph-based models are then passed to a View Correlation Discovery Network (VCDN) [75], which integrates the information to make a final prediction. MOGONet is trained in two phases: in the first phase, only the graph-based models are trained for 500 epochs using the Adam optimizer with a learning rate of $10^{-3}$, while the VCDN is not trained. In the second phase, both the graph-based models and the VCDN are trained for 2500 epochs, with the learning rate for the graph-based models halved ($5 \times 10^{-4}$) and the VCDN trained with a learning rate of $10^{-3}$, using the Adam optimizer. All coding was done on a CPU-enabled Google Colab notebook using PyTorch framework.

**Author contributions**

HAR designed and conceptualized the study. AS wrote the paper. HAR, FH, AA, MY, NHL and AS edited the manuscript. AS carried out all the analyses, including statistical analyses and benchmarking. AS generated all figures and all tables. All authors have read and approved the final version of the paper.

**Conflict of interest**

The authors declare no competing financial and non-financial interests.

**Funding**

This study was supported by the UNSW Scientia Program Fellowship and the Australian Research Council Discovery Early Career Researcher Award (DECRA), under grant DE220101210 to HAR.

**Data and tool availability**

The source code and the dataset can be found at https://github.com/alrzsdgh/Graph-Review/.

**Acknowledgments**

Table 4. Detailed information of the papers reviewed in this study.

| Study | Objective | Task Level | Modalities Used | Graph Construction | Interpretability Objective | Interpretability Technique | Interpretability Mode |
|---|---|---|---|---|---|---|---|
| Yang et al. [19] | Classifying bipolar disorder (manic depression) vs healthy cases | Graph | sMRI and fMRI | **Nodes:** Each patient's brain ROIs **Edges:** Fully connected graph where each edge's weight is calculated using Pearson correlation between node's features. **Feature vector:** Each node's feature vector is generated using combination of features from fMRI and sMRI. | Importance of each modality  Importance of each feature and their coactivation | Replacing modalities' features with dummy variable  Attention within model structure | Post processing  Interpretable Model |
| Keicher et al. [35] | To predict COVID-19 outcome | Node | CT, radiomic, and meta data | **Nodes:** Each patient **Edges:** Distance between nodes are calculated using weighted Minkowski on the combination of radiomic and meta data features and each node is connected to its k nearest neighbors. **Feature vector:** A combination of features from all modalities. | Which neighbor patients have more impact on a specific patient prediction (proper for pandemic situations) | Attention in GAT | Interpretable Model |
| Qu et al. [20] | To predict various physiological phenotypes | Graph | emoid-fMRI and nback-fMRI | For each patient, two graphs based on each modality are constructed. **Nodes:** Brain ROIs. **Edges in each graph:** Pearson correlation between nodes is calculated and the k-nearest neighbor approach implemented subsequently. **Feature vector:** Features from each modality for their corresponding graph. | Importance of each modality  Importance of each node (brain's ROI)  Importance of each edge (relationship between ROIs) | Eliminating modalities  Grad – RAM  Edge masking | Post processing  Post processing  Post processing |
| Wang et al. [36] | Biomedical classification | Node | mRNA expression, DNA methylation, and microRNA expression data | For each patient, three graphs are constructed based on each modality. **Node:** Each patient **Edges in each graph:** Cosine similarity between each node is calculated and edges with value higher a certain threshold are maintained. **Feature vector:** Features from each modality for their corresponding graph. | Importance of each modality | Eliminating modalities | Post processing |

| Study | Objective | Task Level | Modalities Used | Graph Construction | Interpretability Objective | Interpretability Technique | Interpretability Mode |
|---|---|---|---|---|---|---|---|
| Schulte-Sasse [27] | Different cancer gene prediction | Node | Mutations, DNA methylation and gene expression | For each cancer, a separate semi-labeled graph is constructed. **Nodes:** genes **Edges:** gene – gene interactions **Feature vectors:** A combination of features from all modalities. | Importance of each modality | Eliminating modalities | Post processing |
| Sebenius [21] | To detect Schizophrenia patients | Graph | fMRI and Morphometric similarity | For each patient, two graphs are constructed based on each modality **Nodes:** Each patient's brain ROIs **Edges:** Pairwise correlation based on each graph's node features is calculated and edges with values higher than a certain threshold are maintained. **Feature vector:** Features from each modality for their corresponding graph. | Importance of each modality  Importance of each ROI | Eliminating modalities  Top – k ranking | Post processing  Post processing |
| Zhang et al. [28] | To classify breast cancer subtypes | Graph | Protein and gene expression profiles | **Nodes:** Protein **Edges:** Protein – protein interactions **Feature vector:** A combination of features derived from different modalities. | Importance of each feature | SHAP | Post processing |
| Chen et al. [22] | To detect autism spectrum disorder | Graph | sMRI and rsfMRI | **Nodes:** Brain ROIs **Edges:** Fully connected graph where each edge's weight is calculated using Pearson correlation between node's based on their fMRI features. **Feature Vector:** A combination of features from both modalities. | Importance of each modality  Importance of each feature | Eliminating modalities  Gradient saliency map | Post processing  Post processing |
| Pfeifer et al. [29] | Cancer patient classification | Graph | mRNA gene expression and DNA methylation | **Nodes:** Protein **Edges:** Protein – protein interactions **Feature vector:** A combination of features from both modalities. | The most important subgraph in each patient's graph | GNNExplainer | Post processing |

| Study | Objective | Task Level | Modalities Used | Graph Construction | Interpretability Objective | Interpretability Technique | Interpretability Mode |
|---|---|---|---|---|---|---|---|
| Zhou et al. [23] | Alzheimer disease classification | Graph | Structural MRI (VBM-MRI), fluorodeoxyglucose PET (FDG-PET), and 18-F florbetapir PET (AV45-PET) | **Nodes:** Each patient's brain ROIs<br>**Edges:** Gaussian similarity function of Euclidean distances of each node based on their features is calculated and each node is connected to its k nearest neighbors.<br>**Feature vector:** A combination of features derived from each modality. | Importance of each ROI | Graph masking | Post processing |
| Chan et al. [37] | Parkinson disease classification | Node | Image (fMRI, DTI)<br><br>omics (SNP, sncRNA, miRNA, RNA sequencing and DNA Methylation) | Two graphs are constructed based on imagery and omics features<br>**Nodes:** Patients<br>**Edges:** Fully connected with each edge's weight calculated based on Pearson correlation between feature nodes.<br>**Feature vector:** imagery features are fed to its corresponding graph and after processing by GCN, the outputs are fed as feature to omics graphs. | Importance of each modality | Eliminating modalities and attention mechanism | Post processing |
| Safai et al. [24] | Parkinson disease classification | Graph | fMRI, TI weighted MRI, DWI | **Nodes:** Each patient's brain ROIs<br>**Edges:** Conditional probability based on DWI features and considering a certain threshold<br>**Feature vector:** A combination of features derived from all modalities. | Importance of each modality | Eliminating modalities | Post processing |
| | | | | | Importance of each feature | Gradient saliency map | Post processing |
| | | | | | Importance of each ROI | Attention mechanism | Interpretable model |
| Li et al. [38] | To classify breast cancer patients into their subtypes | Node | Copy number variation (CNV) mRNA, and reverse phase protein array data (RPPA) | **Nodes:** Patients<br>**Edges:** Similarity of nodes is calculated based on Euclidean distance and subsequently, k-nearest method is applied to maintain more important edges.<br>**Feature vector:** A combination of features derived from all modalities. | Importance of each modality | Sensitivity analysis | Post processing |

| Study | Objective | Task Level | Modalities Used | Graph Construction | Interpretability Objective | Interpretability Technique | Interpretability Mode |
|---|---|---|---|---|---|---|---|
| Bi et al. [25] | Alzheimer disease classification | Graph | fMRI and SNP | **Nodes:** Each patient's brain ROIs and genes<br>**Edges:** Two types of edges are constructed. First called 'weights' based on Pearson correlation between nodes and considering a certain threshold. The second one is 'similarity-based' which is draw when two nodes share a certain number of neighbors.<br>**Feature vector:** Features from fMRI are used for ROIs and features from SNP are utilized for genes. | Importance of each modality | Eliminating modalities | Post processing |
| | | | | | Importance of each node (brain ROI) and each feature | Gradient saliency map | Post processing |
| Li et al. [34] | To find targets for drugs | Node | Protein and drug | Two graphs are constructed (topology graph and feature graph)<br>**Nodes:** Drug – protein pairs<br>**Edges:** For topology graph, two nodes are linked if they share a drug or protein. For feature graph, cosine similarity between nodes is calculated and each node is connected to its k nearest neighbors.<br>**Feature vector:** A combination of feature derived from proteins and drugs. | Importance of other nodes on each node's prediction | Attention in GAT | Interpretable model |
| Pfeifer et al. [30] | To predict the survival status of patients suffering kidney cancer | Graph | Gene expression (mRNA) and DNA Methylation | One graph for each patient<br>**Nodes:** Proteins<br>**Edges:** Protein-protein interactions<br>**Feature vector:** features derived from each modality | Importance of each modality | Comparing the performance of the decision tree generated by each specific module | Interpretable Model |
| | | | | | Importance of each feature | Gini impurity index (a standard tree-based importance measures) | Interpretable Model |
| | | | | | Importance of each node (protein) | SHAP | Post Processing |

| Study | Objective | Task Level | Modalities Used | Graph Construction | Interpretability Objective | Interpretability Technique | Interpretability Mode |
|---|---|---|---|---|---|---|---|
| Kaczmarek et al. [31] | To classify cancer | Graph | microRNA and mRNA | One graph for each patient **Nodes:** mRNA and miRNA molecules **Edges:** Edges are formed between nodes based on miRNA-mRNA targeting using TargetScan. **Feature vector:** respective expression of the molecule in each patient | Modality importance | modality elimination | Post Processing |
| | | | | | Which nodes are more important to classify each cancer | Graph Transformer | Interpretable Model |
| | | | | | Which mRNA and miRNA interaction plays crucial role in each cancer classification | Graph Transformer | Interpretable Model |
| Kan et al. [26] | To predict patients' gender | Graph | fMRI and DTI | **Nodes:** Each patient's brain ROI **Edges:** Nodes are connected if they are correlated via fMRI (pairwise correlation) or DTI (tractography algorithm) **Feature vector:** features derived from each modality | Importance of each feature Importance of each brain ROI | weights values in the model | Interpretable Model |
| Bintsi et al. [39] | Brain aging prediction | Node | Imaging (structural MRI and diffusion weighted MRI) clinical data | **Nodes:** Patients **Edges:** Weighted features are generated by feeding Imagery and non-imagery features to an MLP. By calculating the probability of connection between two nodes and a modified version of KNN method, edges are draw. **Feature vector:** Feature derived from imagery data | Importance of each modality | Eliminating modalities | Post processing |
| | | | | | Importance of each feature | Attention mechanism | Interpretable model |

| Study | Objective | Task Level | Modalities Used | Graph Construction | Interpretability Objective | Interpretability Technique | Interpretability Mode |
|---|---|---|---|---|---|---|---|
| Kazi et al. [40] | To detect patients with Alzheimer's disease<br><br>To predict patient's age and sex | Node | Cognitive tests, MRI ROIs measures, PET imaging, DTI ROI measures, demographics, etc.<br><br>MRI and fMRI images | Automatic graph construction through training process<br>**Nodes:** Patients<br>**Edges:** edge values are calculated using Euclidean distance between nodes based on their features and edges with values higher than a threshold (defined through training process) are maintained.<br>**Feature vector:** Features derived from mentioned modalities for each task | Feature importance | Attention mechanism | Interpretable Model |
| Tang et al. [44] | To predict 30-day all-cause hospital readmission rate | Node | EHR data and chest radiograph | Two graphs are constructed for each modality<br>**Nodes:** each node is a hospital admission<br>**Edges:** for both graph, edges are calculated based on the Euclidean distance between node's EHR features and values bellow a threshold are eliminated<br>**Feature vector:** features extracted from each modality | Importance of each feature and neighboring nodes | GNNExplainer | Post processing |
| Bi et al. [41] | To detect Alzheimer's disease | Node | fMRI and SNP genetic data | **Nodes:** Patients<br>**Edges:** two nodes remain connected if their connection strength ranks among the top k edge values calculated based on the Euclidean distance of fMRI or SNP data. Then the remained edge values are calculated based on Pearson correlation between connected nodes.<br>**Feature vector:** concatenation of the features derived from each modality | Importance of each feature | graph convolutional filter weights | Interpretable Model |

| Study | Objective | Task Level | Modalities Used | Graph Construction | Interpretability Objective | Interpretability Technique | Interpretability Mode |
|---|---|---|---|---|---|---|---|
| Huo et al. [32] | Cancer survival prediction | Graph | Pathological slide, clinical records and genomic profile | One graph is generated for each modality **Nodes:** image sections in pathological graph, demographic and medical reports for clinical graph, and five genomic embeddings using GSEA for genomic graph **Edges:** clinical and genomic graphs are fully connected while nodes in pathological graph are connected to their adjacent spatial neighbors **Feature vector:** features derived by feeding pathological slides to KimiaNet for pathological graph, one-hot encodings for clinical graph, and genomic embeddings for genomic graph | Importance of different regions in pathological slides and features of clinical records  Importance of each feature in genomic profile | Attention mechanism  Integrated gradient analysis | Interpretable Model  Post processing |
| Xiao et al. [42] | To predict sarcopenia | Node | Demographics and Lab data | **Nodes:** Patients **Edges:** patient similarity is calculated based on the demographic's features (cosine similarity) Lab features (DTW). These similarities are then combined. Each node is connected to its m neighbors with highest edge value. **Feature vector:** features derived from each modality | Contribution of each patient on other patient's prediction | Attention in GAT | Interpretable Model |
| Ouyang et al. [43] | Disease classification | Node | mRNA expression, DNA methylation and miRNA expression | for each modality, a graph is constructed dynamically through model training **Nodes:** Patients **Edges:** are calculated based on weighted cosine similarity considering node's features **Feature vector:** for each graph, features derived from each modality | Modality importance  Feature importance | Attention mechanism  Inner product regularization | Interpretable Model  Interpretable Model |
| Ma et al. [33] | Learns relations among cells and genes | Node | Genes, proteins, or peak regions | **Nodes:** Genes and cells **Edges:** a gene node is connected to a cell node if the gene exists in the cell **Feature vector:** features derived from each modality | Importance of genes to a specific cell | Attention mechanism | Interpretable Model |